\documentclass[12pt]{article}
\usepackage{graphicx}
\usepackage{amsmath,amssymb,epsfig,bm}

\newcommand\tr{\mathop{\mathrm{tr}}}

\newcommand{\be}{\begin{eqnarray}}
\newcommand{\ee}{\end{eqnarray}}

\def\lsim{\mathrel{\rlap{\lower3pt\hbox{\hskip1pt$\sim$}}
     \raise1pt\hbox{$<$}}} 
\def\gsim{\mathrel{\rlap{\lower3pt\hbox{\hskip1pt$\sim$}}
     \raise1pt\hbox{$>$}}} 

\def\la{\langle}\def\ra{\rangle}
\def\del{\partial}

\def\bi{\bibitem}
\def\tr{{\rm tr}}

\begin{document}

\begin{titlepage}
\begin{flushright}
\end{flushright}
\vskip 1.7in
\begin{center}
{\bf\Large {Integrating Holographic Vector Dominance \\
to Hidden Local Symmetry \\ for the Nucleon Form Factor }}
\vskip 1.5cm
{Masayasu Harada}
\vskip 0.05in
{\small{ \textit{Department of Physics,  Nagoya University,  Nagoya,
      464-8602, Japan}}}
\vskip 0.6cm
{Mannque Rho}
\vskip 0.05in
{\small{\textit{Institut de Physique Th\'eorique,  CEA Saclay, 91191
      Gif-sur-Yvette C\'edex, France \&\\
Department of Physics, Hanyang University, Seoul 133-791, Korea}}}

\end{center}
\vskip 0.7in
\baselineskip 16pt
\date{\today}

\begin{abstract}
We derive a two-parameter formula for the electromagnetic form factors
of the nucleon described as an instanton by ``integrating out" all KK
modes other than the lowest mesons from the infinite-tower of vector
mesons in holographic QCD while preserving hidden local symmetry for
the resultant vector fields. With only two parameters, the proton
Sachs form factors can be fit surprisingly well to the available
experimental data for momentum transfers $Q^2\lsim 0.5$ GeV$^2$ with
$\chi^2$/dof $\lsim 2$. We interpret this agreement as indicating the
importance of an infinite tower
in the soliton structure of the nucleon. The prediction of the
Sakai-Sugimoto holographic dual model is checked against the fit
values to assess its accuracy in describing the proton structure. We
find that the structure of the ``core" of roughly 1/3 in the proton
size indicated in experiments and commonly associated with an intrinsic
quark-gluon structure in QCD is ``hidden" in the infinite tower in the holographic model.
\end{abstract}

\end{titlepage}

\section{Introduction}
\label{sec:intro}
The celebrated Sakurai vector dominance (sVD for short) model for the
EM form factors~\cite{sakurai} works surprisingly well for the
mesons~\cite{sakurai-vd} but it famously fails for the
nucleon~\cite{IJL,BRW,iachello,bijker}\footnote{As a measure of
  ``goodness" and ``badness" of the sVD, the $\chi^2$/dof for the
  electric form factor up to momentum transfer $Q^2\sim 1$ GeV$^2$ is
  4.3 and 1116, respectively for the pion and the nucleon.}. The
failure has been interpreted as an indication that the nucleon has a
``core" which is not present in the pion structure~\cite{IJL}. The
``core" has been attributed -- among a variety of possibilities -- to
a compact microscopic structure of QCD variables, such as for instance
a little chiral bag with quarks confined within~\cite{BRW}.

The recent development of holographic dual QCD (hQCD for short),
specially, the Sakai-Sugimoto string theory
model~\cite{sakai-sugimoto} that implements correctly chiral symmetry
of QCD, indicates a dramatic return of the notion of vector dominance
for {\em both} mesons and
baryons~\cite{HRYY,HRYY:VD,Hashimoto,Kim:2008pw}. What characterizes
the baryon structure in the hQCD model is that the baryon emerges as a
soliton in the presence of an infinite tower of vector mesons. It is
the infinite tower that renders the VD description applicable to both
mesons and baryons. In fact, the isovector component of the EM form
factor has the ``universal" form\footnote{As for the nucleon form
  factor, we are referring here specifically to the isovector
  component of the charge form factor. However the same discussion
  applies to both the Sachs electric and magnetic form factors
  measured in experiments as we will see later.}
\be
F_V^h=\sum_{k=0}^\infty
 \frac{g_{v^{(k)}}g_{v^{(k)}hh}}{Q^2+m_{2k+1}^2}\label{inftower}
\ee
where $g_{v^{(k)}}$ is the photon-$v^{(k)}$ coupling
and $g_{v^{(k)}hh}$ is the
$v^{(k)}$-$hh$ coupling for $h=\pi, N$. Here $v^{(k)}$ is the $k$-th
isosvector vector mesons $\rho, \rho^\prime, ...$ in the tower. What
distinguishes the hadron probed is the vector-meson coupling to the
hadron. This form -- which is almost completely saturated by the first
four vector mesons -- works surprisingly well for both mesons and
baryons at low-momentum transfers, say, $Q^2\lsim 0.5$ GeV$^2$ as we
shall show below.

Three questions arise from these results.

The first is what makes the difference between the ``good" sVD for the
pion and the ``bad" sVD for the nucleon disappear when the infinite
tower is present?
For this issue, it is perhaps important to note that even for the pion
form factor, the sVD is quite fragile under certain external
conditions. For instance, in hidden local symmetry
theory~\cite{BKUYY:BKY,HY:PR}
that will be the main tool of this paper, the sVD of the pion form
factor is shown to break down in thermal
background~\cite{VD-violation} and is expected to break down even more
precociously in dense matter.

The second is the problem of the ``core." While it is reasonable to
view the pion as point-like when probed at long wave length, the
nucleon is an extended object, for which a local field approximation
must break down at some not too high momentum scale. There are
indications from high-energy proton-proton scattering, experiments on
high mass muon pairs and also in deep inelastic scattering off nucleon
that the nucleon has a core of $\sim 0.2 - 0.3$ fm in
size~\cite{core}. It is this class of observations that led to the
notion of the ``Little Bag"~\cite{BRlittlebag} and to the hybrid
structure of the nucleon with a core made up of a quark bag surrounded
by a meson cloud~\cite{BRW}. The question is whether the core is lodged
in the infinite tower and if so, in what form? Closely tied to this
question is: Does the photon ``see" the size of the instanton -- which
is a skyrmion in the infinite tower of vector mesons -- which goes as
$\sim 1/\sqrt{\lambda}$ where $\lambda=N_c g_{YM}^2$ is the 't Hooft
constant ? For a large value of $\lambda$, the soliton (instanton)
size is small. However this size cannot be a physical quantity. The
physical size should be independent of $\lambda$. The instanton size
therefore must be akin to the bag size which is also unphysical
according to the Cheshire Cat principle. So the last question is: Is
the ``core" a physical observable?

The objective of this paper is to address the above three questions.
\section{Holographic Form Factors}
Let us first define the notations that we shall use and then give a
concise summary of the hQCD calculation of the form factors as
described in \cite{HRYY,HRYY:VD,Hashimoto}. We shall follow the
notations of \cite{HRYY,HRYY:VD}.

The nucleon form factors are defined from the matrix elements of the
external currents,
\begin{equation}
\la p^{\prime}|J^{\mu}(x)|p\ra=e^{iqx}\,\bar
u(p^{\prime})\,{\cal O}^{\mu}(p,p^{\prime})\,u(p) \:,
\end{equation}
where $q=p^{\prime}-p$ and $u(p)$ is the nucleon spinor of momentum
$p$.
From the Lorentz invariance and the current conservation, with the
assumption of CP invariance, the operator ${\cal O}^{\mu}$ takes the
form
\begin{equation}
{\cal
O}^{\mu}(p,p^{\prime})
=\gamma^{\mu}\frac12\left[F_1^s(Q^2)+F^a_1(Q^2)\tau^a\right]
+i\frac{\sigma^{\mu\nu}}{2m_N}q_{\nu}\frac12
\left[F_2^s(Q^2)+F^a_2(Q^2)\tau^a\right]\:,
\end{equation}
where $m_N\simeq 940~{\rm MeV}$ is the nucleon mass,
$F_1^s$ and $F_2^s$ are the Dirac and Pauli form factors
for isoscalar current respectively, and  $F_1^a$, $F_2^a$ are for
isovector currents.

As matrix elements, the form factors contain all one-particle
irreducible diagrams
for two nucleons and one external current $A_\mu$ given as
\begin{equation}
\la p^{\prime}|J^{\mu}(x)| p \ra=\la p^{\prime}|
\frac{\delta}{\delta A_{\mu}} e^{iS_{\rm eff}[A]}|p\ra\ .
\end{equation}
Thus they are very difficult to calculate in QCD. It turns out,
however, that the anti-de
Sitter/conformal field theory (AdS/CFT) correspondence, or
gravity/gauge theory correspondence, found in
certain types of string theory, enables one to compute such
non-perturbative quantities as hadron form factors within certain
approximations.

According to the AdS/CFT correspondence, the low energy effective action of the gravity
dual of QCD becomes the generating functional for the correlators
of an operator ${\cal O}$ in QCD in the large $N_c$ limit, i.e., $
e^{iS_{5D}^{\rm eff}[\phi(\epsilon,x)]}=\left<\exp\left[i\int_x
\phi_0\,{\cal O}\right]\right>_{\!\!\rm QCD}$ where $\phi(z,x)$ is a bulk field, acting as a source for ${\cal O}$
when evaluated at the UV boundary $z=\epsilon$. Now the normalizable modes of the bulk field are identified as the physical states
in QCD, created by the operator ${\cal O}$.

The model we shall use is the gravity dual of low energy QCD  with
massless flavors in the large $N_c$ (or quenched) approximation
constructed by Sakai and Sugimoto
(SS)~\cite{sakai-sugimoto}.\footnote{We shall use the notation and
  numerical values of Ref.\cite{HRYY}. Bulk baryon fields are introduced to represent the soliton configurations but the description with them is equivalent to collective-quantizing solitons as is done in \cite{Hashimoto}.}
The holographic dual of spin-$1\over 2$ baryons, or nucleons, in this
model with two flavors was constructed in \cite{HRYY} by introducing a
bulk baryon field, whose effective action is given in the ``conformal
coordinate" $(x,w)$ as
\begin{eqnarray}
S_{5D}^{\rm eff}=\int_{x,w}\left[-i\bar{\cal B}\gamma^m D_m {\cal B}
-i m_b(w)\bar{\cal B}{\cal B} +\kappa(w)
\bar{\cal B}\gamma^{mn}F_{mn}^{SU(2)_I}{\cal B}+\cdots \right]
+S_{\rm meson},
\label{5dfermion1}
\end{eqnarray}
where ${\cal B}$ is the 5D bulk baryon field, $D_m$ is the gauge
covariant derivative with $m=1,2,3,4,5$,
$\gamma^{mn}$ is defined as
$\gamma^{mn} = (1/2) \left[ \gamma^m\,,\,\gamma^n\right]$
and $S_{\rm meson}$ is the effective action for the mesons. $\kappa (w)$ is an effective warped constant that depends on the holographic coordinate $w$. There is
an additional parity-odd term called Chern-Simons term,  which is not
specified since it is not needed for our discussion. Using the
instanton nature of baryon, the coefficients $m_b(w)$ and $\kappa(0)$
can be reliably calculated in string theory. In (\ref{5dfermion1}) the
ellipsis stands for higher derivative terms -- higher in
$\alpha^\prime$ or in $1/\lambda$ -- that
are expected to be suppressed at low energy, $E<M_{KK}$, where the KK
mass sets the cut-off scale. Note that the magnetic coupling involves
only the non-abelian part of the flavor symmetry $SU(2)_I$ -- with
abelian $U(1)_B$ being absent -- due to the non-abelian nature of
instanton-baryons.

In computing the electromagnetic (EM) form factors for the nucleons
using the AdS/CFT correspondence,
we need to identify the dual bulk field of the external EM current,
which is the bulk photon field. Since the electric charge operator is
the sum of the isospin and the baryon charge, $Q_{\rm em}=I_3+\frac12
B$, we have to identify a combination of $A_{\mu}^3$ and $A_{\mu}^B$,
the third component of the isospin gauge field and the $U(1)_B$ gauge
field, respectively, as the photon field.
Then all baryon bilinear operators in the effective action
that couple to either $U(1)_B$ gauge fields or $SU(2)_I$ gauge
fields will contribute to the EM form factors.

The (nonnormalizable) photon field is written as
\begin{equation}
A_{\mu}(x,w)=\int_q\,A_{\mu}(q)A(q,w)\,e^{iqx}\:,
\end{equation}
with boundary conditions
that $A(q,w)=1$ and $\partial_wA(q,w)=0$ at the UV boundary,
$w=\pm w_{\rm max}$ and
the (normalizable) bulk baryon field as
\begin{equation}
{\cal B}(w,x)=\int_p\left[f_L(w)u_L(p)+f_R(w)u_R(p)\right]e^{ipx}\,.
\end{equation}
These 5D wave functions, $A(q,w)$ and $f_{L,R}(w)$, are determined by
solving the equation of motion from our action (\ref{5dfermion1}).
Then, using the AdS/CFT correspondence, one can read off the form
factors.
The Dirac form factor
$F_1(Q^2)=F_{1{\rm min}}\,Q_{\rm em}+F_{1{\rm mag}}\,I_3$
with ($Q^2\equiv -q^2$) is of the form
\begin{eqnarray}
F_{1{\rm min}}(Q^2)&=&\int_{-w_{max}}^{w_{max}} dw\,\left|f_L(w)\right|^2\,A(q,w)\:,\label{ff_1}\\
F_{1\rm mag}(Q^2)&=&2\int_{-w_{max}}^{w_{max}}
dw \kappa(w)\left|f_L(w)\right|^2\partial_w A(q,w)\:,\label{ff_1m}
\end{eqnarray}
where $F_{1{\rm min}}$ is from the minimal coupling, and $F_{1\rm mag}$ the magnetic coupling.
Similarly the Pauli form factor is given as $F_2(Q^2)=F^3_{2}(Q^2)\,I_3$ with
\begin{eqnarray}
F^3_{2}(Q^2)&=&4\,m_N\,
\int_{-w_{max}}^{w_{max}} dw\,
\kappa(w)f_L^*(w)f_R(w)\,A(q,w)\:,
\label{ff_2}\end{eqnarray}
which comes solely from the magnetic coupling.
The form factors (\ref{ff_1}), (\ref{ff_1m}) and (\ref{ff_2}) receive corrections
from the higher order operators in the effective action~(\ref{5dfermion1}), but they are expected to be suppressed by powers of  $E/M_{KK}$ at low energy. Note however that our result contains {\em full} quantum effects in the large $N_c$ limit.

As shown in \cite{HRYY}, one can replace the form factors by
an infinite sum of vector-meson exchanges~\cite{HRYY}, if we expand the nonnormalizable photon field
in terms of the normalizable vector meson $\psi_{(2k+1)}$ of mass $m_{2k+1}$ as
\begin{equation}
A(q,w)=\sum_{k=0}^\infty \frac{g_{v^{(k)}}\psi_{(2k+1)}(w)}{Q^2+m_{2k+1}^2}\:,
\label{expansion}
\end{equation}
where the decay constant of the $k$-th vector mesons  is given as
$g_{v^{(k)}}=m_{2k+1}^2\zeta_k$ with
\begin{equation}
\zeta_k=\frac{\lambda N_c}{108\pi^3}M_{KK}\int_{-w_{max}}^{w_{max}}
 dw\frac{U(w)}{U_{KK}}\,\psi_{(2k+1)}(w) \:,
\label{def:zeta}
\end{equation}
where $U_{KK}$ is a parameter of the SS model and
\begin{equation}
dw=\frac32\frac{U_{KK}^{1/2}}{M_{KK}}\frac{dU}{\sqrt{U^3-U_{KK}^3}}\,.
\end{equation}
The resulting EM form factors then take the
form
\begin{eqnarray}
F_{1}(Q^2)&=&F_{1{\rm min}}\,Q_{\rm em}+F_{1{\rm mag}}\,\frac{\tau^3}{2}=
\sum_{k=0}^{\infty} \left(g^{(k)}_{V,min}Q_{\rm em}
+g_{V,mag}^{(k)}\,\frac{\tau^3}{2}\right)\frac{\zeta_k m_{2k+1}^2}
{Q^2+m_{2k+1}^2}\,,\label{F1}\\
F_2(Q^2) &=&F_2^3(Q^2)\,\frac{\tau^3}{2}= \,\frac{\tau^3}{2}\sum_{k=0}^{\infty}
{g_2^{(k)}\zeta_k m_{2k+1}^2\over Q^2+m_{2k+1}^2}\,,\label{F2}
\end{eqnarray}
where
\begin{eqnarray}
g_{V,min}^{(k)}&=&\int_{-w_{max}}^{w_{max}} dw\,\left|f_L(w)\right|^2
\psi_{(2k+1)}(w)\label{gvmin}\\
g_{V,mag}^{(k)}&=& 2\int_{-w_{max}}^{w_{max}}
 dw \,\kappa(w)\left|f_L(w)\right|^2
 \partial_w \psi_{(2k+1)}(w)\:,\label{gvmag}\\
 g_2^{(k)}&=&
 4m_N \int_{-w_{max}}^{w_{max}}
dw \,\kappa(w) f_L^*(w)f_R(w) \psi_{(2k+1)}(w)\label{g2k}\:.
\end{eqnarray}
This is the set of formulas that we shall use for our analysis that follows.

\section{Integrating Out the Tower}
In order to focus on the role that the lowest vector mesons $V^{(0)}\equiv
(\rho,\omega)$ play as in the sVD, we would like to integrate out all
other vector mesons than the $V^{(0)}$. The resulting form factors will be
given in terms of the properties of the $V^{(0)}$ arranged in power series
of the momentum transfer involved, $Q^2$. Since the resulting hidden
local theory is endowed with a chiral invariance (lightly broken by
the quark masses) as discussed in \cite{HY:PR}, we would like to do
this to the next-to-leading order (NLO) in the expansion. This means
that the form factors will involve terms up to ${\cal O}(p^4)$ in the
HLS Lagrangian. We will follow \cite{Harada:2006di,Harada:2010iv} in
deriving the formulae for the proton. We will restrict to the number
of flavors to be 2 and focus on the vector channel only, i.e., $U(2)$
symmetric $\rho$ (isovector) and $\omega$ (isoscalar). Unless required
otherwise, we will show only the $\rho$ contribution with the
appropriate $\omega$ contribution understood.

An aspect which appears to play an important role in understanding the
infinite-tower structure of the form factors is that the local vector
meson fields that figure in the tower in the bulk sector are degrees
of freedom in a warped space with certain geometry. What we will do is
to integrate out all the vector degrees of freedom lying above the
$M_{KK}$ scale, which leaves only the $V^{(0)}\equiv (\rho,\omega)$. We
suppose that this corresponds in some sense to integrating out the
$\rho^\prime (1450), \rho^{\prime\prime} (1700),...$ and
$\omega^\prime (1420), \omega^{\prime\prime} (1650), ...$ in the dual
gauge sector. This procedure may be considered as a part of the
renormalization group flow in the radial direction (namely the fifth
direction $z$) in the bulk sector which may be associated with the
renormalization group flow of the boundary (gauge) theory. Because of
the warped geometry in the bulk sector, it is not clear that one can
simply map the $v^{(k)}$'s that are integrated out in the bulk sector to the $\rho^k$'s
integrated out in the gauge sector. What we will do below is to
integrate out the $v^{(k)}$'s for $k>0$ at the classical level but in a
certain warped geometrical background and it is possible that this
does not necessarily map one-to-one to integrating out all $\rho$'s in
the gauge sector lying higher than $\Lambda_\chi\sim 1$ GeV. Thus the
two-parameter formula derived here may not be directly compared with
the generalized vector dominance models employed  in the literature to
accurately fit the nucleon form factor data. This aspect will be
discussed in Section \ref{core}.

\subsection{\it Hidden local symmetry (HLS)}

Following the strategy proposed in \cite{Harada:2010iv} for the pion
form factor, we wish to integrate out all KK modes other than $V^{(0)}=(\rho,
\omega)$ such that the resulting action is hidden local symmetric in
$V^{(0)}$. The reason for resorting to HLS is to keep track of chiral
symmetry in making power expansion~\cite{HY:PR}. Now in doing this,
the equations of motion for the higher KK modes are used to replace
them in favor of the $V^{(0)}$ to ${\cal O}(p^4)$ in the derivative
expansion. Going to higher order may not be justified unless one
incorporates higher order terms in $\alpha^\prime$ of the DBI action
(i.e., higher $1/\lambda$ terms) and loop corrections (i.e., higher
$1/N_c$ terms) in the bulk sector, the task of which seems at present
out of reach of systematic treatments. We are essentially
``integrating out" the tower at the tree level with higher tower
effects lodged in the action given to ${\cal O}(p^4)$. It is in this
sense that we are exposing the infinite tower effect as {\em
  corrections} to the lowest KK mode. The validity of such a procedure
is clearly limited to low momentum transfer. We are thus limiting our
consideration to $Q^2\lsim 0.5$ GeV$^2$. We should also note that
confining to $N_f=2$, we are leaving out certain other degrees of
freedom (such as $\phi$ that enters in the isoscalar channel in the
gauge sector).

We start with a brief recap of the method used in \cite{Harada:2010iv}
for the pion form factor which is immediately applicable to the
nucleon form factors as described in Appendix.

Consider the meson action $S_{\rm meson}$ in (\ref{5dfermion1}) of the
Sakai-Sugimoto hQCD model~\cite{sakai-sugimoto} in the form
compacified in ~\cite{HRYY}\footnote{Our convention is
\begin{equation}
\mbox{tr}\,\big[ T_a T_b \big] = \frac{1}{2} \delta_{ab}
\ ,
\quad
(a,\,b) = 0, 1, 2, \ldots, N_f^2-1\ ,
\end{equation}
with
\begin{equation}
{\cal A}_\mu = {\cal A}_\mu^a T_a \ .
\end{equation}}
\be
S_{\rm meson}=- \int d^4 x  dw {1\over 2 e^2(w)} \;
 \mbox{tr}\, F_{mn}F^{mn}\label{SS}
\ee
where $m=(\mu, z) = (1,2,3,4,5=z)$.
Taking the ${\cal A}_5=0$ gauge, the 5D gauge field ${\cal A}_\mu$ is
expanded as~\footnote{
  Here we follow the normalization of the wave function
  $\psi_{2k+1}(w)$ adopted in Ref.~\cite{HRYY}, which is
  different from the one in Ref.~\cite{Harada:2010iv}.
  Note that the vector meson fields in Ref.~\cite{HRYY}, say
  $\rho_\mu^{(k)}$, is related to the fields $V_\mu^{(k)}$
  as $V_\mu^{(k)} = g_{2k+1} \rho_\mu^{(k)}$ with $g_{2k+1}$
  being the HLS gauge couplings.
This $g_{2k+1}$ is related to the parameter $\zeta_k$ given
in Eq.~(\ref{def:zeta}) as $\zeta_k =1/ g_{2k+1}$.
}
\begin{eqnarray}
  {\cal A}_\mu(x, w)
 &=& \alpha_{\mu ||} (x)
  +
  \hat{\alpha}_{\perp\mu}(x) \psi_0(w)
\nonumber \\
&&
+
\sum_{m=1}^\infty
A_\mu^{(m)}(x) \psi_{(2m)}(w)
-
\sum_{k=0}^\infty
\, \hat{\alpha}_{\parallel \mu}^{(k)}(x)
\,\zeta_k \, \psi_{(2k+1)}(w)
\, ,
\label{Amuxz:2}
\end{eqnarray}
where
\begin{equation}
\hat{\alpha}^{(k)}_{\parallel \mu}(x) =
\alpha_{\mu ||}(x) - V_\mu^{(k)}(x)
\end{equation}
and
\be
\alpha_{\mu ||}(x) &=&
\frac{1}{2i} \left[
    \partial_\mu \xi_R \, \xi_R^\dag
    + i \xi_R {\mathcal R}_\mu \xi_R^\dag
    + \partial_\mu \xi_L \, \xi_L^\dag
    + i \xi_L {\mathcal L}_\mu \xi_L^\dag
\right]
\ ,\\
\hat{\alpha}_{\perp\mu}(x)&=&{\alpha}_{\perp\mu}(x)= \frac{1}{2i} \left[
    \partial_\mu \xi_R \, \xi_R^\dag
    + i \xi_R {\mathcal R}_\mu \xi_R^\dag
    - (\partial_\mu \xi_L \, \xi_L^\dag
    + i \xi_L {\mathcal L}_\mu \xi_L^\dag)
\right]
\
\ee
Here $A_\mu^{(n)}$ and $V_\mu^{(n)}$ are, respectively, 4D axial vector and vector fields at $n$-th tower.
An important observation exploited in \cite{Harada:2010iv} is that in the ${\cal A}_5=0$ gauge, there is the residual gauge symmetry
\be
{\cal A}_\mu (x,z)\to h(x)(i\del_\mu +{\cal A}_\mu (x,z))h^\dagger (x)
\ee
with $h\in SU(2)_F$. This means that when KK-reduced from 5D to 4D,
${\cal A}_\mu$ will be given as a stack of infinite tower hidden local
gauge fields~\cite{sakai-sugimoto}. With the constraint
\be
\phi^R (z)+\phi^L (z)-\sum_{k=0}^\infty \zeta_k \psi_{(2k+1)} (z)=1
\ee
the transformations are
\begin{eqnarray}
V_\mu^{(k)} (x)&\rightarrow&
  h(x)(i\del_\mu+V_\mu^{(k)} (x))h^\dagger (x)\\
A_\mu^{(m)} (x)&\rightarrow&
  h(x)(A_\mu^{(m)} (x))h^\dagger (x)\\
\alpha_{\mu_\parallel} (x)&\rightarrow&
  h(x)(i\del_\mu+\alpha_{\mu_\parallel})h^\dagger (x)\\
\alpha_{\mu \perp} (x)&\rightarrow&
  h(x)(\alpha_{\mu \perp} (x))h^\dagger (x).
\end{eqnarray}
We would like to integrate out {\em all} axial-vector fields {\em as
  well as} the tower of the vector fields except the lowest, $V^{(0)}$, such
that the hidden gauge invariance for the $V^{(0)}$ be preserved to the NLO
in the action (\ref{SS}). This can be done by setting
\be
A_\mu^{(m)} (x)&=& 0\ \ {\rm for\ all}\ \ m\,,\\
\alpha_{\mu ||}(x) - V_\mu^{(k)}(x)&=& 0 \ \ {\rm for} \ k>0.
\ee
These are the equations of motion for the fields to be integrated out
  to the leading chiral order that we will be considering. Note that
  the kinetic energy terms are of higher order so they do not figure.
Now when the resulting gauge field
\begin{eqnarray}
  {\cal A}_\mu(x, w)  &=& \alpha_{\mu ||} (x)
  +
  {\alpha}_{\perp\mu}(x) \psi_0(w)
-
\hat{\alpha}_{\parallel \mu}(x) \,\zeta_0\,\psi_{1}(w)
\, ,
\label{Amuxz:22}
\end{eqnarray}
is substituted into the action (\ref{SS}), one obtains the standard
HLS Lagrangian in 4D,
\be
{\cal L}_{HLS}=&& F_\pi^2 {\tr}[\hat{\alpha}_{\perp\mu }\hat{\alpha}^{\mu \perp}]+ aF_\pi^2 {\tr}[\hat{\alpha}_{\mu ||}\hat{\alpha}^{\mu ||}]\nonumber\\
&&- \frac{1}{2g^2}{\tr}[V_{\mu\nu}V^{\mu\nu}] + {\cal L}_4\label{HLS}
\ee
with
\begin{equation}
\hat{\alpha}_{\mu ||} (x)=\alpha_{\mu ||} (x)- V^{(0)}_\mu (x)\ .
\end{equation}
Here the parameters $F_\pi$ and $a$ are given in terms of the
parameters of the holographic model, namely, $N_c$, $\lambda=g_{YM}^2
N_c$ and $M_{KK}$. ${\cal L}_4$ is the ${\cal O}(p^4)$ Lagrangian in
the chiral expansion in the leading order in $N_c$ that consists of 26
terms with the external vector and axial vector (electroweak) fields
included. We won't write them down here, referring to
\cite{Harada:2006di,Harada:2010iv,Harada:2010cn} for details.
We note that they are all determined
by the three parameters-- $N_c$, $\lambda$, $M_{KK}$ -- and the wave
functions $\psi_0$ and $\psi_1$ for the pion and $V^{(0)}$
respectively. Once the meson sector is fixed, there are no free
parameters for the baryon sector.

Calculating in tree order -- which is the best we can do given that
loops cannot be computed in the bulk sector --, the pion EM form
factor will have the standard form $(1-\frac{a}{2}) +
\frac{a}{2}\frac{m_\rho^2}{Q^2+m_\rho^2}$ plus the contribution from
the ${\cal L}_4$ term. Thus
\be
F_V^\pi=(1-\frac{\tilde{a}}{2})
+ \frac{\tilde{a}}{2}\frac{m_\rho^2}{Q^2+m_\rho^2}
+\tilde{z}\frac{Q^2}{m_\rho^2}+\cdots\label{2parahls}
\ee
with
\be
\tilde{a}=a+\delta a
\ee
where $\delta a$ and $\tilde{z}$ are contributions from the ${\cal
  L}_4$ term. They are entirely determined once the three parameters
are fixed. The ellipsis stands for terms of higher order in $Q^2$.

It is important to note that this is a generic formula that would
arise from {\em any} holographic model that gives 5D YM action in a
warped space. Thus it can come from top-down as from the SS model or
bottom-up as in dimensionally deconstructed models~\cite{son}. Physics
will be encoded in the parameters that will figure in the action. For
instance, the standard Sakurai vector dominance would correspond to
$\tilde{a}=2$ and $\tilde{z}=0$. This point will be important for
later discussions. In HLS theory as an effective theory of QCD,
$\delta a$ and $\tilde{z}$ should receive contributions from one-loop
terms as well as from the ${\cal O}(p^4)$ counter terms. To the
leading order, we would have $\delta a=\tilde{z}=0$, so for the mass
formula $m_\rho^2=2F_\pi^2 g^2$ with $a=2$, we would have the Sakurai
VD formula with the ``contact" term vanishing. In this holographic
model, the ${\cal O}(p^4)$ contributions come, in some RG flow sense
but involving no loops, from the high tower at the classical level.

Given that the formula (\ref{2parahls}) is generic, it makes sense to
consider the $\tilde{a}$ and $\tilde{z}$ as free parameters and fit to
low-momentum data. Such a fit was made in \cite{Harada:2010cn} to
momentum transfers $Q^2\sim 1$ GeV$^2$, finding the best fit parameter
$\tilde{a}=2.44$ and $\tilde{z}=0.08$ with $\chi^2/dof=1.6$. Compare
this to the Sakurai VD with $\tilde{a}=2$ and $\tilde{z}=0$ that gives
$\chi^2/dof=4.3$. Although $Q^2\sim 1$ GeV$^2$ is a bit too high a
momentum transfer for the validity of the expansion, it indicates that
the Sakurai VD has some room for improvement even for the pion
although the deviation is not significant.

The key observation that we will exploit for the baryon is that the
formula (\ref{2parahls}) can be derived directly from the
infinite-tower formula (\ref{inftower}). It was shown in
\cite{Harada:2010cn} that using the sum rules satisfied by $F_V^\pi
(0)$ and $\frac{dF_V^\pi (Q^2)}{dQ^2}|_{Q^2=0}$, the infinite-tower
formula (\ref{inftower}) yields
\be
\frac{\tilde{a}}{2} &=& \frac{g_{\rho}g_{\rho\pi\pi}}{m_{\rho}^2},\\
\frac{\tilde{z}}{m_{\rho}^2}&=&
 -\sum_{k=1}^\infty \frac{g_{v^{(k)}}g_{v^{(k)}\pi\pi}}{m_{2k+1}^4}.
\label{eq:40}
\ee
These quantities are given in terms of the known wave functions
$\psi_0$, $\psi_1$ and their eigenvalues. From the Sakai-Sugimoto
work~\cite{sakai-sugimoto}, it comes out that $\tilde{a}\simeq 2.62$
and $\tilde{z}\simeq 0.08$ with the $\chi^2/dof=2.8$. It is surprising
that the $\chi^2$ comes out better with the SS model than with the Sakurai
VD.

\subsection{\it Two-parameter formulae for the nucleon}

As alluded above, the nucleon form factors can also be given in the
same two-parameter form (\ref{2parahls}). In deriving it, there are
two different options in dealing with the baryons. One option is to
first integrate out all vector mesons in the tower except for the
$V^{(0)}$ from the SS action obtaining the HLS action -- which is just
(\ref{HLS}), construct the soliton baryon from that reduced action and
then compute the form factors. This will be equivalent to the
construction of the holographic baryons discussed in \cite{nawa}. We
will comment on the structure of the form factors obtained in this way
in Section \ref{CBM}. The option we will take in this work is to first
construct the instanton baryon in 5D, and then compute the form
factors with the vector mesons and pions coupled to the instanton
baryon~\footnote{The power of this approach was shown also in
  few-nucleon systems where many-body nuclear interactions are
  involved: The structure of few-nucleon systems comes out to be much
  cloer to experiments than the other method~\cite{sutcliffe}.}. The
physics is captured by the action (\ref{5dfermion1}), which is
completely equivalent to what is obtained by the collective
quantization of Hashimoto et al.~\cite{Hashimoto}.

We shall now derive the two parameter formula from by integrating out
the tower of vectors except for $V^{(0)}$ {\em and} all axial
vectors from the infinite-tower formulae (\ref{F1}) and (\ref{F2}) for
the nucleon. To see how the integrating-out procedure gives rise to an
identical formula for both the pion and the nucleon, it is instructive
to see how the generic structure of the vector dominance
(\ref{inftower}) arises for both of them with the only difference
being in the $v^{(k)} hh$ coupling $g_{v^{(k)} hh}$. It also illustrates the
universal nature of the role of the degrees of freedom that are
integrated out. To illustrate this, we use a slightly different,
simpler and transparent notation since the axial fields are not
involved. From \cite{sakai-sugimoto}, one can write down the direct
$\gamma\pi\pi$ coupling in the form
\begin{equation}
 {\cal L}_{\gamma\pi\pi} \sim
  \left(1-\sum_{k=0}^\infty
\frac{g_{v^{(k)}}g_{v^{(k)}\pi\pi}}{m_{2k+1}^2}\right)
{\tr} ([\pi,\del^\mu\pi]{\cal V}_\mu)
\label{eq:41}
\end{equation}
where ${\cal V}_\mu$ is the photon field
and with \cite{HRYY}, the direct $\gamma NN$ coupling in the form
\be
{\cal L}_{\gamma NN}\sim
\left(1 -\sum_{k=0}^\infty
\frac{g_{v^{(k)}}g_{v^{(k)} NN}}{m_{2k+1}^2}\right)
\bar{B}\gamma^\mu B{\cal V}_\mu.
\ee
Now the charge sum rule gives for the pion and the proton,
\begin{eqnarray}
\sum_{k=0}^\infty \frac{g_{v^{(k)}}g_{v^{(k)}\pi\pi}}{m_{2k+1}^2}&=&1\, , \\
\sum_{k=0}^\infty \frac{g_{v^{(k)}}g_{v^{(k)} NN}}{m_{2k+1}^2}&=&1.
\label{eq:44}
\end{eqnarray}
Thus the direct photon coupling vanishes, i.e.,
${\cal L}_{\gamma\pi\pi}={\cal L}_{\gamma NN}=0$. This implies that
when
the tower of the vector mesons is integrated out leaving only the
$\rho$ meson, a direct photon coupling to the pion and the baryon of
the same form
will reappear in the form factor from the degrees of freedom that are
integrated out. This would signal a deviation from the Sakurai VD.

It will be shown in Appendix how the resulting two-parameter formulae
showing the direct photon coupling can be derived starting from the
action (\ref{5dfermion1}).

For comparison with experiments and for physical interpretation, it is
more convenient to work with the Sachs form factors given in terms of
the Dirac and Pauli form factors given above as
\begin{eqnarray}
G_E^p(Q^2) &=& F_1^p(Q^2) - \frac{Q^2}{4 m_N^2} \, F_2^p(Q^2)\label{GE}
\ ,
\\
G_M^p(Q^2) &=& F_1^p(Q^2) + F_2^p(Q^2).\label{GM}
\end{eqnarray}
The hQCD predictions~\cite{HRYY,HRYY:VD,Hashimoto} are
\begin{eqnarray}
F_1^p(Q^2) &=& F_{\rm 1,min}(Q^2)
  + \frac{1}{2} F_{\rm 1,mag}(Q^2)
\ ,
\nonumber\\
F_2^p(Q^2) &=& \frac{1}{2}\,\sum_{k=0}^{\infty}
  \frac{g_2^{(k)} \, g_{v^{(k)}}}
   {Q^2 + m_{2k+1}^2}
\end{eqnarray}
with
\begin{eqnarray}
F_{\rm 1,min}(Q^2) &=& \sum_{k=0}^{\infty}
  \frac{g_{V,{\rm min}}^{(k)} \, g_{v^{(k)}}}
   {Q^2 + m_{2k+1}^2}
\ ,
\nonumber\\
F_{\rm 1,mag}(Q^2) &=& \sum_{k=0}^{\infty}
  \frac{g_{V,{\rm mag}}^{(k)} \, g_{v^{(k)}}}
   {Q^2 + m_{2k+1}^2}
\ .
\end{eqnarray}
The Sachs form factors have simple physical interpretations as the
spatial Fourier transforms of the charge and magnetization
distributions of the nucleon in the Breit frame.
Note that since we are dealing with the EM current, we have replaced
$\zeta_k \, m_{2k+1}^2$ by the photon-$v^{(k)}$ coupling
$g_{v^{(k)}}$, i.e.,
\be
g_{v^{(k)}}=\zeta_k \, m_{2k+1}^2\ .
\ee
Now following the procedure worked out for the pion form factor, when
higher KK modes other than the lowest one, $V^{(0)}$, are integrated
out,
we can immediately write down the resultant Sachs form factors in the
generic form
\be
G_E^p (Q^2)
&=& \left( 1 - \frac{a_E}{2} \right) + z_E\, \frac{Q^2}{m_V^2}
 + \frac{a_E}{2} \, \frac{m_V^2}{m_V^2 + Q^2 }
\ ,
\label{GE form}
\\
G_M^p (Q^2) / \mu_p
&=& \left( 1 - \frac{a_M}{2} \right) + z_M\, \frac{Q^2}{m_V^2}
 + \frac{a_M}{2} \, \frac{m_V^2}{m_V^2 + Q^2 }
\ ,
\label{GMform}
\ee
where $a_E$, $z_E$, $a_M$ and $z_M$ are parameters to be determined,
and we have written $m_1=m_V$ with $V$ standing for $V^{(0)}$ and
assumed that $m_\rho=m_\omega = m_V$.\footnote{Hereafter we will write $V$ for $V^{(0)}$ unless otherwise noted.}
Now formally integrating out higher modes is equivalent to expanding
(\ref{GE form}) and (\ref{GMform}) in $Q^2/m_{2k+1}^2$ for $k \geq 1$
up to ${\mathcal O}(Q^2)$ while keeping the $\rho$ ($\omega$) meson
($k=0$ mode)
propagator as it is.

In what follows we show explicitly the expressions for the electric
form factor but the same procedure holds for the magnetic form
factor.

The result for the electric form factor takes the form
\begin{eqnarray}
G_E^p(Q^2)
&= &
  \left(
    g_{V,{\rm min}}^{(0)} + \frac{1}{2} g_{V,{\rm mag}}^{(0)}
  \right)
  \frac{\zeta_0 \, m_{V}^2 }
   {Q^2 + m_{V}^2}
\nonumber\\
&&
  {} +
  \sum_{k=1}^{\infty}
  \left(
    g_{V,{\rm min}}^{(k)} + \frac{1}{2} g_{V,{\rm mag}}^{(k)}
  \right)
  \frac{\zeta_k \, m_{2k+1}^2 }{m_{2k+1}^2}
  \left[ 1 - \frac{Q^2}{m_{2k+1}^2} \right]
\nonumber\\
&&
 {} - \frac{Q^2}{8 m_N^2} \sum_{k=0}^{\infty}
  \frac{g_2^{(k)} \, \zeta_k \, m_{2k+1}^2 }{m_{2k+1}^2}
\ .
\end{eqnarray}
Now thanks to the sum rules
\begin{eqnarray}
\sum_{k=0}^{\infty}
  \left(
    g_{V,{\rm min}}^{(k)} + \frac{1}{2} g_{V,{\rm mag}}^{(k)}
  \right)\,\zeta_k
 = 1
\ ,
\nonumber\\
\sum_{k=0}^{\infty}
  g_2^{(k)} \, \zeta_k = g_2
\ ,
\label{eq:53}
\end{eqnarray}
the form factor is reduced to
\begin{eqnarray}
G_E^p(Q^2)
&= &
  \left(
    g_{V,{\rm min}}^{(0)} + \frac{1}{2} g_{V,{\rm mag}}^{(0)}
  \right)
  \frac{\zeta_0 \, m_{V}^2 }
   {Q^2 + m_{V}^2}
\nonumber\\
&&
  {} +
  \left[ 1 -
    \left(
      g_{V,{\rm min}}^{(0)} + \frac{1}{2} g_{V,{\rm mag}}^{(0)}
    \right)
    \zeta_0
  \right]
  - \frac{Q^2}{m_{V}^2}\,
  \sum_{k=1}^{\infty}
  \left(
    g_{V,{\rm min}}^{(k)} + \frac{1}{2} g_{V,{\rm mag}}^{(k)}
  \right)
  \frac{\zeta_k \, m_{V}^2 }{m_{2k+1}^2}
\nonumber\\
&&
 {} - \frac{Q^2}{8 m_N^2} g_2
\ \label{GE-expand}
\end{eqnarray}
Comparing (\ref{GE form}) and (\ref{GE-expand}),
we find the $a$ and $z$ parameters for the electric form factor as
\begin{eqnarray}
a_E^{\rm (hQCD)} &=& 2
    \left(
      g_{V,{\rm min}}^{(0)} + \frac{1}{2} g_{V,{\rm mag}}^{(0)}
    \right)
    \zeta_0
\ ,
\\
z_E^{\rm (hQCD)} &=& -
  \sum_{k=1}^{\infty}
  \left(
    g_{V,{\rm min}}^{(k)} + \frac{1}{2} g_{V,{\rm mag}}^{(k)}
  \right)
  \frac{\zeta_k \, m_{V}^2 }{m_{2k+1}^2}
  - \frac{m_{V}^2}{8 m_N^2} \, g_2 \ .
\label{z hQCD}
\end{eqnarray}
Making a similar expansion for the magnetic form factor, one finds from (\ref{GMform})
\begin{eqnarray}
a_M^{\rm (hQCD)} &=&
 \left[ 2
    \left(
      g_{V,{\rm min}}^{(0)} + \frac{1}{2} g_{V,{\rm mag}}^{(0)}
    \right)
    \zeta_0
  + g_2^{(0)} \zeta_0
 \right]
 / \mu_p
\ ,
\\
z_M^{\rm (hQCD)} &=& -
\sum_{k=1}^{\infty}
\left[
  \left(
    g_{V,{\rm min}}^{(k)} + \frac{1}{2} g_{V,{\rm mag}}^{(k)}
  \right)
  \zeta_k
  + \frac{1}{2} g_2^{(k)} \zeta_k
\right]
\frac{m_{V}^2 }{m_{2k+1}^2}
/ \mu_p
\ ,
\label{z m hQCD}
\end{eqnarray}
with $\mu_p = 1 + (1/2) g_2$.

\section{Numerical Analysis}

The form factor we will analyze has the generic form characterized by
two parameters $a$ and $z$
\be
G(Q^2)/\beta
= \left( 1 - \frac{a}{2} \right) + z\, \frac{Q^2}{m_V^2}
 + \frac{a}{2} \, \frac{m_V^2}{m_V^2 + Q^2 }.\label{gen}
\label{eq:59}
 \ee
This form holds for the pion form factor with $\beta=1$ and the nucleon
form factors with $\beta=1$ for the electric and $\mu_p$ for the
magnetic form factor (assuming $m_\rho=m_\omega$). Hadron structure
enters in the $a$ and $z$ parameters.
\subsection{\it Best Fit}
 We first consider seriously the structure of the form factor
 (\ref{gen}) given at low momentum transfer, say, $Q^2\ll M_{KK}^2$ or
 (in QCD) $Q^2\ll \Lambda_\chi^2$. We could then subject the
 two-parameter formula to experimental data. In this spirit, we
 best-fit (\ref{gen}) to the accurate experimental data given, say, in
 Ref.~\cite{Arrington:2007ux}. Given that the approximation is valid
 for low momentum transfers, we limit to $Q^2\leq 0.5$
 GeV$^2$. Throughout we use the values
\begin{equation}
m_V = 0.775 \,\mbox{GeV}
\ ,
m_N = 0.938 \, \mbox{GeV}
\ .
\label{eq:60}
\end{equation}

\vskip 1.5cm
\begin{figure}[btp]
\begin{center}
\includegraphics[width=7.5cm]{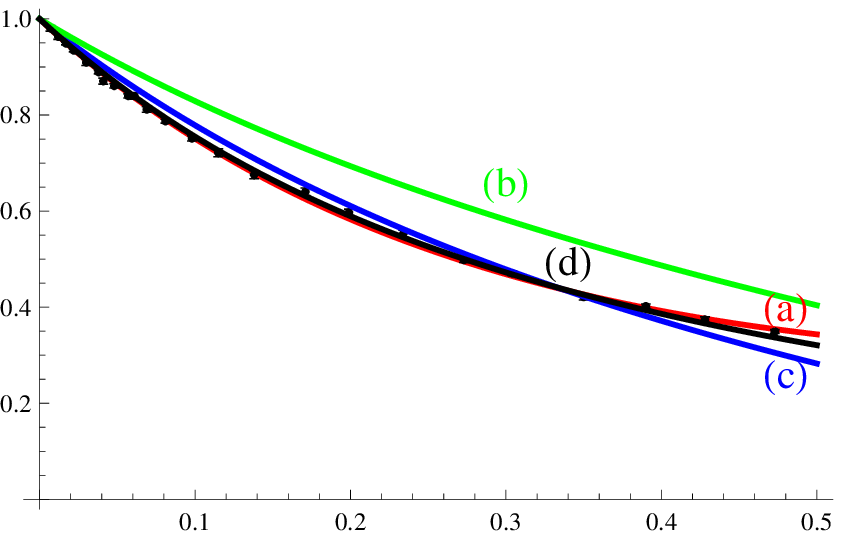}
\includegraphics[width=7.5cm]{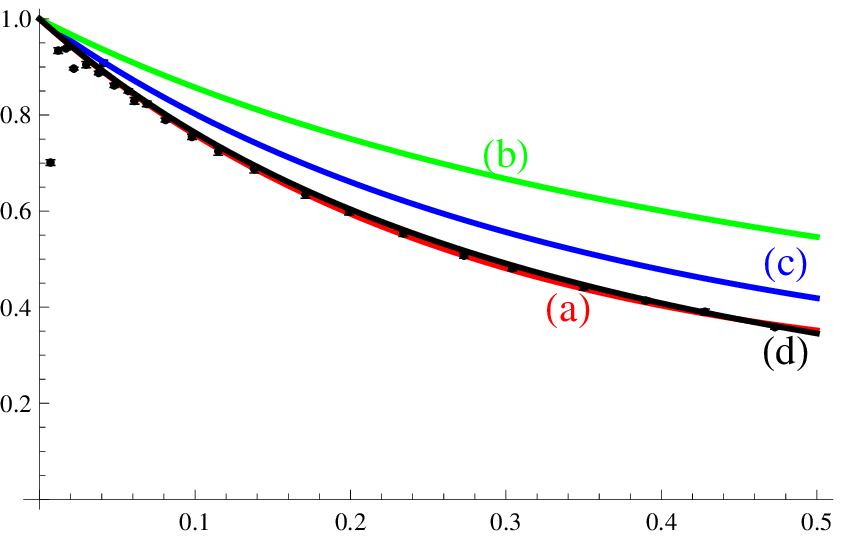}
\end{center}
\caption[]{(Color online) $G_E^p$ (left panel) and $G_M^p$ (right panel) vs. $Q^2$. Vertical axis shows the value of
$G_{E,M}^p$, while the horizontal axis shows the value of $Q^2$ in
unit of $\mbox{GeV}^2$. The red curve (a) is the ``best fit," the green curve (b) the sVD prediction, the blue curve (c) the hVD prediction and the black curve (d) the Bijker-Iachello two-component model prediction described below.
}\label{fig1}
\end{figure}

\begin{figure}[bhtp]
\begin{center}
\includegraphics[width=7.5cm]{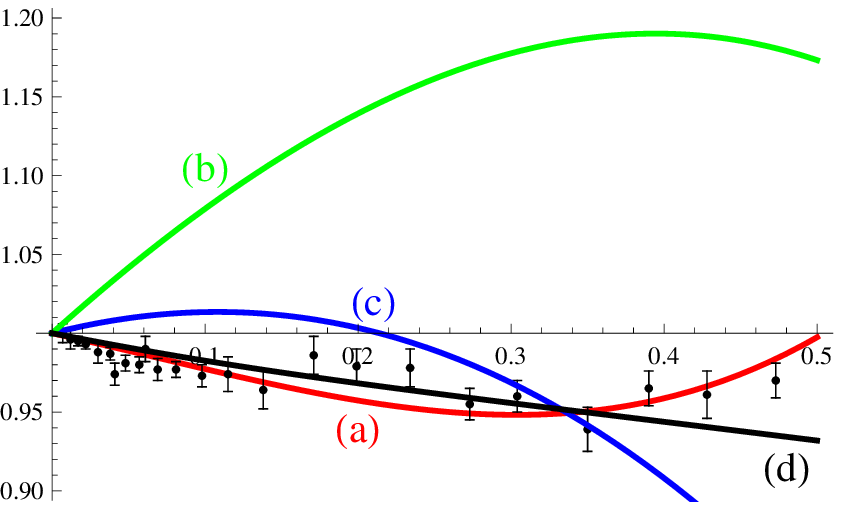}
\includegraphics[width=7.5cm]{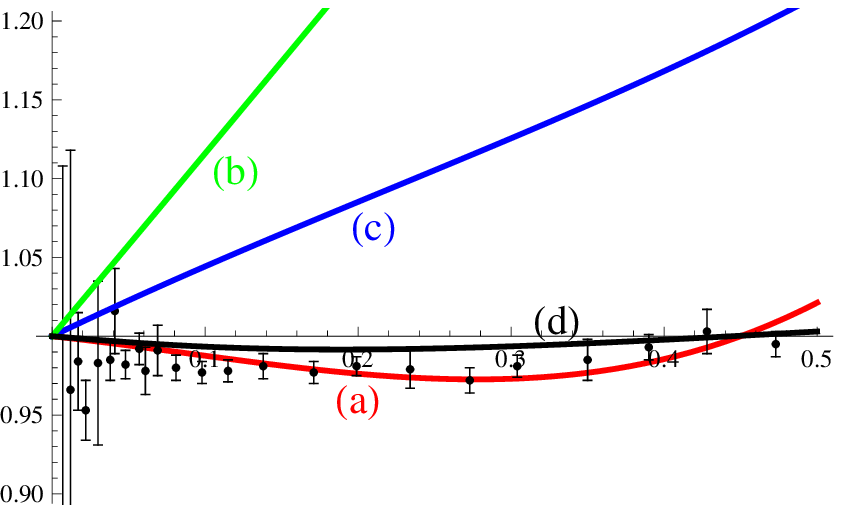}
\end{center}
\caption[]{(Color online) $G_E^p/G_D$ (left panel) and $G_M^p/\mu_D G_D$ (right panel) vs. $Q^2$. Vertical axis shows the value of
$G_{E,M}^p/G_D$, while the horizontal axis shows the value of $Q^2$ in unit of $\mbox{GeV}^2$. The red (a), green (b), blue (c) and black (d) curves are for, respectively, the ``best fit," the sVD prediction, the hVD prediction and the Bijker-Iachello two-component model fit.}\label{fig2}
\end{figure}

\begin{figure}[tbhp]
\begin{center}
\includegraphics[width=10cm]{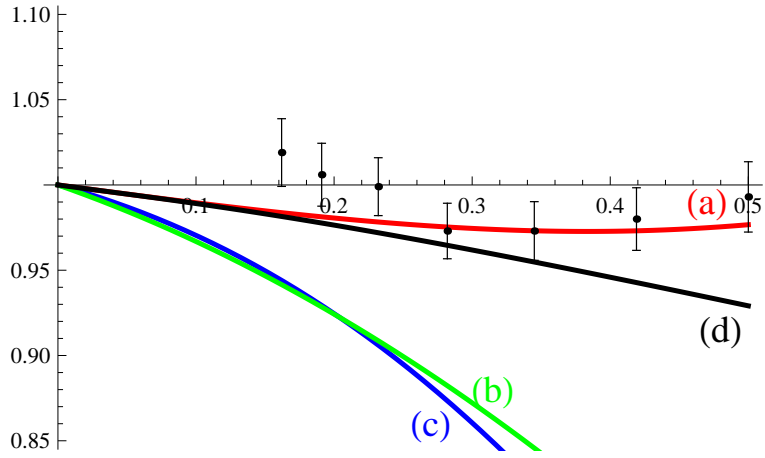}
\end{center}
\caption[]{(Color online) $\mu_p G_E^p/G_M^p$ vs. $Q^2$ given by the ``best fit"(a, red),  sVD (b, green), hVD (c, blue) and BI two-component model (d, black) compared with the experiments of \cite{ratioGE-GM}.}\label{ratioGE-GM}
\end{figure}

The best $\chi^2$ fit is given in red (labeled as (a)) in Fig \ref{fig1} for the form factors and in Fig \ref{fig2} for the same divided by the dipole form factor
\be
G_D(Q^2)=\left(\frac{1}{1+\frac{Q^2}{0.71\ {\rm GeV}^2}}\right)^2.
\ee
The best fit parameters and $\chi^2$'s are:
\be
G_E^p:&& a_E^{(best)}=4.55, z_E^{(best)}=0.45; \chi_E^2/{\rm dof}=1.5\label{bestGE}\\
G_M^p:&& a_M^{(best)}=4.31, z_M^{(best)}=0.40; \chi_M^2/{\rm dof}=1.1.\label{bestGM}
\ee
It is surprising that the fit is so good with so few parameters. A more stringent test is the ratio $\mu_p G_E^p/G_M^p$ for which fairly accurate data are available at low momentum transfers~\cite{ratioGE-GM}. The predicted ratios are plotted and compared with the experiments in Fig.~\ref{ratioGE-GM}. The agreement of the fit parameters (\ref{bestGE}) and (\ref{bestGM}) with the experiments is surpring, considerably better than other models discussed below.

Since the fit parameters are close to each other with similar $\chi^2$ -- although we are aware of no reason why they should be, it is tempting to take $a_E=a_M$ and $z_E=z_M$ and make the best-fit. The result is: $a^{(best)} =  4.42$ and $z^{(best)} =  0.42$ with $\chi^2$/dof = 1.90. It is interesting to compare the best-fit for the nucleon to the best-fit for the pion obtained in \cite{Harada:2010cn} using the ``universal" formula (\ref{gen}): $a_\pi^{(best)}=2.44$ and $z_\pi^{(best)}=0.08$ with $\chi^2$/dof=2.44 while the sVD with $a=2$ and $z=0$ gives $\chi^2$/dof=4.3. The low-momentum data are more accurate for the nucleon than the pion and that accounts for the better $\chi^2$ of the nucleon form factors. It is interesting to note that in the case of the pion, the deviation in $\chi^2$ of the best-fit from the sVD is relatively small accounting for the general acceptance of the sVD.


\subsection{\it sVD and hVD predictions}\label{hqcd}
As noted, it is the parameters $(a,z)$ that carry information on the nucleon structure. We now examine how the vector dominance models, Sakurai vector dominance (sVD) and holographic vector dominance (hVD), fare in predicting these parameters and in fitting the data for the nucleon.

First we consider the sVD. It has been known that the sVD does not work at all contrary the case for mesons. The sVD corresponds to taking\footnote{Note that unlike $z_M^{VD}$, $z_E^{VD}$ differs from 0 because of the second term in the expression for $G_E$, eq.~(\ref{GE}).}
\be
a_E^{VD} &=&2, z_E^{VD}=-m_\rho^2/(4 m_N^2)\simeq -0.171, \label{VDE}\\
a_M^{VD} &=&2, z_M^{VD}=0.
\ee
The result is shown in green (labeled as (b)) in Figs.~\ref{fig1} and \ref{fig2}. The $\chi^2$/dof comes out to be 187 and 852, respectively, for $G_E^p$ and for $G_M^p$. This reconfirms the well-known story that sVD simply fails for the nucleon.

Now turning to hVD, we will use the results of \cite{HRYY} given in Table~\ref{table2} reproduced from \cite{HRYY}.
\begin{table}[hbp]
\begin{center}
\begin{tabular}{|c|c|c|c|c|c|c|c|}
\hline $k$ & $m^2_{2k+1}$ & $\zeta_k$
& $g^{(k)}_{V,min}$ & $g_{V,mag}^{(k)}$ & $ g_{V,min}^{(k)}\zeta_k$
& $ g_{V,mag}^{(k)}\zeta_k$   & $g_2^{(k)}\zeta_k$  \\
 \hline\hline
0& 0.67 & 0.272 & 5.933 & -0.816 & 1.615 & -0.222 & 3.323  \\
 \hline
1 & 2.87 & -0.274 & 3.224 & -1.988 & -0.882& 0.544 & -1.918 \\
 \hline
2 & 6.59 & 0.272  & 1.261 & -1.932 & 0.343 & -0.526 & 0.828 \\
\hline
3 & 11.8 & -0.271 & 0.311 & -0.969 & -0.084 & 0.262 & -0.243 \\
\hline \hline {\rm sum} & - &- & -& -& 0.992 & 0.058  &1.989($g_2=2.028$)\\
\hline
\end{tabular}
\end{center}
\caption{\small  Numerical results for vector meson couplings for
the lowest four excitations in the case $\lambda N_c=50$ (taken from \cite{HRYY}). Sum
rules hold to a high precision. Our convention for the vector
meson fields differ by sign from that of Sakai and Sugimoto for
odd $k$. The vector meson mass squared is in the unit of
$M_{KK}^2$.} \label{table2}
\end{table}
We are limiting to the lowest four states since it is found numerically that the charge and magnetic sum rules are almost completely saturated by them~\cite{HRYY:VD}. One should however be careful in using this observation for form factors since the four states may not saturate momentum-dependent observables as fully as the static quantities.

By using the values listed in Table~\ref{table2}, the parameter $a$ comes out to be
\begin{eqnarray}
a_E^{\rm (hQCD)} = 3.01 \ ,
\label{a v hQCD}
\\
a_M^{\rm (hQCD)} = 3.14 \ ,
\label{a m v hQCD}
\end{eqnarray}
As for $z_{E,M}$, there are no known sum rules for the sums in Eqs.~(\ref{z hQCD}) and (\ref{z m hQCD}). We shall simply take the values for $k=1,2,3$ from Table \ref{table2}. We find
\begin{eqnarray}
z_E^{\rm (hQCD)} &\simeq& -
  \sum_{k=1}^{3}
  \left(
    g_{V,{\rm min}}^{(k)} + \frac{1}{2} g_{V,{\rm mag}}^{(k)}
  \right)
  \frac{\zeta_k \, m_{V}^2 }{m_{2k+1}^2}
  - \frac{m_{V}^2}{8 m_N^2} \, g_2
\nonumber\\
&=&
-0.042 \ ,
\label{z v hQCD}
\\
z_M^{\rm (hQCD)} &\simeq& -
\sum_{k=1}^{3}
\left[
  \left(
    g_{V,{\rm min}}^{(k)} + \frac{1}{2} g_{V,{\rm mag}}^{(k)}
  \right)
  \zeta_k
  + \frac{1}{2} g_2^{(k)} \zeta_k
\right]
\frac{m_{V}^2 }{m_{2k+1}^2}
/ \mu_p
\nonumber\\
&=&
0.16
\ .
\label{z m v hQCD}
\end{eqnarray}

The form factors predicted parameter-free in hQCD given by
\be
(a_E,z_E)&=&(3.01, -0.042)\nonumber\\
(a_M,z_M)&=&(3.14, 0.16),\nonumber
\ee
are plotted in blue (labeled as (c)) in Figs.~~\ref{fig1} and \ref{fig2}. The $\chi^2$ comes out to be 20.2 for $G_E^p$ and 133 for $G_M^p$. While $G_E$ comes out to be rather reasonable, $G_M$ is less so, although vastly better than with sVD. We will speculate how this comes about in the discussion section.

It is important for later discussions to note that in both the best fit and the SS holographic model, the direct photon coupling represented by $D=1-a/2$ is negative. This is in a stark contrast to the result $D\geq 0$ obtained in all QCD-motivated models as will be elaborated below.
\section{Comparison with Other Models}
It is interesting to compare what we have found in the holographic
model with those found in other descriptions. In doing so, we discover
that there is a basic difference from the results of chiral
perturbation theory, generalized vector dominance model (GVDM) and
chiral bag model (CBM). This points to a puzzle on the role of the
``core" in the proton structure mentioned in Introduction.

The puzzle we face will surely be resolved by future lattice QCD
calculations. At present, it is not. There are lattice QCD results at
momentum transfers up to $Q^2\sim 1.4$ GeV$^2$. With the minimum quark
mass reached at $m_\pi\sim 300$ MeV, however, the form factors are
found to scale less slowly than the experimental and their fit is
markedly less good than hQCD~\cite{lattice}.

At the next level of fundamental approach, there have been a large
number of calculations in baryon chiral perturbation theory involving
nucleons and pions, i.e., baryon chiral perturbation theory
(BChPT). Calculations to ${\cal O}(p^4)$ in BChPT provide a decent
description up to only $Q^2\sim 0.1$ GeV$^2$ and fail for higher
$Q^2$. For a reasonable description, it seems indispensable to go to
${\cal O}(p^5)$ involving two-loops or alternatively implement
explicit vector-meson degrees of freedom. With the vector mesons
suitably introduced, it has been possible to extend the treatment with
certain success to momentum transfers $Q^2\sim 0.4$
GeV$^2$~\cite{gegelia}. Since chiral perturbation theories, with or
without vector mesons, deal with the local baryon field, there is
always non-vanishing point photon coupling, and the numerical
importance of vector mesons can only be accidental.

Much more pertinent to our considerations are the GVDM and CBM to
which we turn.
\subsection{\it GVDM}
As mentioned in Introduction, the failure of sVD in describing the
nucleon form factors has led to numerous efforts to modify the vector
dominance structure. The most successful approach to improve the fits
is purely phenomenological in nature. It consists of bringing in more
massive vector mesons with widths, if available, and coupling to
continuum with the asymptotic $Q^2$ behavior of perturbative QCD,
thereby enabling one to go to higher momentum transfers $Q^2 >1$
GeV$^2$. One such sophisticated model is the Lomon model that consists
of $\rho$ (with its width), $\rho^\prime (1.45)$, $\omega$ and
$\omega^\prime (1419)$ and hadron form factors including the
logarithmic momentum transfer behavior of asymptotic
freedom~\cite{lomon}. This model is found to be highly successful in
representing the existing high-quality data up to $Q^2\sim 10$
GeV$^2$.

For our purpose, it suffices to consider the two-component model of Bijker and Iachello~\cite{iachello,bijker} (BI for short). It captures the essential features that we are interested in. This model with six parameters -- which improves on
the Iachello-Jackson-Lande model~\cite{IJL} -- illustrates both long- and short-distance structures we would like to unravel. In the form used therein, it attempts, by implementing the widths and perturbative QCD effects, a global fit to nucleon form factors up to $Q^2\sim 10$ GeV$^2$. The BI fit is plotted in black (labeled as (d)) in Figs~\ref{fig1} and \ref{fig2}. The corresponding $\chi^2$ up to $Q^2=0.5$ GeV$^2$ comes to be $\chi^2/dof =
2.0$ and $6.1$ respectively for $G^p_{E,M}$. The fit is quite good, but despite the larger number of parameters involved, it is not as good as the two-parameter fit with (\ref{bestGE}) and (\ref{bestGM}).

This two-component model allows one to give a meaning to, and make a simple discussion on, the ``core." The ``core" size obtained in \cite{bijker} is $\sim (0.3-0.4)$ fm, comparable to what is observed in nature.  To see how this comes about in this model, we will drastically simplify the BI parametrization. For this we ignore $\phi$ which contributes to the isoscalar form factor and the widths of the vector mesons, both of which are absent in the hQCD formulation. The first is because we are working with two flavors and the second because they are higher order in $1/N_c$. We also assume the flavor $U(2)$ symmetry so $m_\rho=m_\omega$. Then $G_E^p$ of \cite{bijker} can be written in the form (the same argument holds for $G_M^p$)\footnote{Since (\ref{bi}) is simplified and approximated from the multi-parameter BI formula by retaining only three parameters, one cannot expect it to do as well as the six-parameter fit. Indeed the $\chi^2$/dof comes out 17.6 which is comparable to that of the SS model.}
\be
G_E^p(Q^2)\approx
g(Q^2)\left((1-\frac{a_{BI}}{2})
+z_{BI}\frac{Q^2}{m_\rho^2}+\frac{a_{BI}}{2}
\frac{m_\rho^2}{m_\rho^2+Q^2}\right).\label{bi}
\label{eq:70}
\ee
Here $g(Q^2)$ stands for the asymptotic $Q^2$ behavior of
perturbative QCD parameterized in \cite{iachello,bijker} as
$g(Q^2)=(1+\gamma Q^2)^{-2}$. The fit parameters of BI that yield the BI results in Figs~\ref{fig1} and \ref{fig2} give
\be
a_{BI}\approx 1.64, \ \ z_{BI}\approx -0.23\label{BIvalue}
\ee
with $\gamma=0.515$ GeV$^{-2}$.

In the two-component model of \cite{bijker}, the core contribution to the form factor is identified to be the {\it deviation from the vector dominance}, namely, $g(Q^2)\left((1-\frac{a_{BI}}{2}) +z_{BI}\frac{Q^2}{m_\rho^2}\right)$ in (\ref{bi}). This gives the size
\be
\la r^2\ra_{core}=-\frac{6}{m_\rho^2}\left(z_{BI}-2\gamma (1-\frac{a_{BI}}{2}) m_\rho^2\right)\approx (0.36\ {\rm fm})^2.
\ee
This reproduces the core size obtained in \cite{bijker}, which is to be compared with the core measured in the experiments~\cite{core}, i.e., $\sim (30-40)\%$ of the proton charge radius $\sqrt{\la r^2\ra_{E,p}}\simeq (0.875\ {\rm fm})$. This feature with a varying core size is generally reproduced in {\em all} sophisticated phenomenological models of the Lomon type~\cite{lomon}. As we will discuss below, the relevant quantity for the BI model is the sign of $D\equiv 1-a_{BI}/2$ which controls the sign and magnitude of the core size. It will turn out that the two-parameter formula we derived from the infinite tower structure of hQCD exhibits a qualitatively different feature.
\subsection{\it CBM}\label{CBM}
One way closer to the quark/gluon degrees of freedom of QCD that
removes the difficulty of the sVD for the nucleon form factors was suggested in
\cite{BRW} in terms of the chiral bag model of the nucleon. It anchors
on the role of pion in the baryon structure dictated by the
spontaneously broken chiral symmetry of QCD. The basic idea is that a
small confined region of quarks and gluons, namely, ``bag," is
surrounded by the meson cloud of pions and the photon couples both
directly to the bag, i.e., an intrinsic core, and through the meson
cloud. This idea is realized by the chiral bag that consists of a
quark-bag coupled to meson cloud via chiral boundary condition. The
simplest picture suggested in \cite{BRW} is the ``little bag" coupled
at the bag surface to the chiral field at the magic chiral angle
$\theta=\pi/2$ at which half of the baryon charge is lodged inside the
bag and the other half outside. Imagine the photon coupling to the
proton via a ``bag" with a coefficient $a$ and through a vector meson
with a coefficient $b$. Now treat the bag as a point-like object. The
charge conservation requires of course that $a+b=1$. On the other
hand, the bag does not contribute to the anomalous magnetic moment
whereas the coupling through the $\rho$ vector meson contributes, via
the tensor $\rho$-$NN$ coupling, $eb\frac{\kappa_\rho}{2m_N}$. So the
anomalous magnetic moment of the proton will be given by
\be
\kappa_V=b\kappa_\rho.
\ee
Thus the chiral bag at the magic angle $a=b=1/2$ will give
\be
\kappa_\rho\approx 2\kappa_V.\label{tensorcoupling}
\ee
Experimentally $\kappa_\rho=6.6\pm 0.6$ and $\kappa_V=\mu_p
-\mu_n=3.71$, so (\ref{tensorcoupling}) is more or less satisfied. The
magic angle is just an idealization and it is possible that $b$ could
be greater than 1/2.

This idealized half-and-half model works fairly well for the proton
charge form factor up to $Q^2\sim 0.4$ GeV$^2$ as one can see in
Fig.~4 of \cite{BRW}. In this picture, there is a core provided by the
bag carrying half of the proton charge resembling the two-component
model described above with a comparable size. A similar description arises when the bag is
replaced by a Skyrme soliton. Suppose that the skyrmion has a size of
the baryon as in the standard skyrmion (i.e., the Skyrme soliton) or
the skyrmion in the presence of the $\rho$ meson in the hQCD
Lagrangian where all other mesons than the $\rho$ have been integrated
out from the tower as in \cite{nawa}. Such a skyrmion will carry all
the charge of the proton and hence should describe -- modulo small
contributions from the fluctuating pion field -- most of the form
factor structure. It turns out that such a description does not fare
well in comparison with nature and in fact, demands incorporating
fluctuating vector meson ($\rho$ and $\omega$ fields) coupled to the
soliton for better description. It has in fact been recognized since sometime that the vector mesons could play an important role in the electromagnetic structure of the skyrmion model for the nucleon~\cite{meissner}. Indeed the resulting structure is analogous
to the chiral bag description described above. It has been shown that
this ``$\pi,\rho,\omega$" model, with mild adjustment of parameters,
can actually describe very well all nucleon form factors to $Q^2\sim 10$
GeV$^2$~\cite{holzwarth}. In some sense there is a manifestation here
of the Cheshire Cat property~\cite{cheshire} where the bag and soliton
structures could be traded in.

The bottom line of all these structures in the models anchored on the
premise of QCD is that they all predict the existence of a ``core" to
which photon couples point-like.

\subsection{\it The problem of the ``core"}\label{core}

The spatial Fourier transform of the form factors $G_E^p(Q^2)$ ($G_M^p
(Q^2)$) describes the charge (magnetic moment) distribution of the proton,
so the two-parameter formula should reflect how the charge (and
magnetic moment) is distributed spatially.  The question we ask is: Can one say about the ``core" from the two-parameter best-fit result and the SS model?.

To see what it is all about, we need to specify what we mean by ``core." As mentioned above, in the two-component model of \cite{bijker} and also the chiral bag model of \cite{BRW}, one may define the core of, say, the proton to be the (extended) component of the proton that is not accounted for by the vector-meson cloud.  In the BI model, what is significant is that the constant term in (\ref{bi}) is not only positive, $g(Q^2)(1-\frac{a_{BI}}{2})>0$ but also $1-\frac{a_{BI}}{2}\lsim 1/2$. This feature is shared by the chiral bag model at the magic chiral angle $\pi/2$ with a half-and-half sharing of the baryon charge.

Now the best-fit case with (\ref{bestGE}) (and also with (\ref{bestGM})) is drastically different. Here because of the fact that the lowest-vector-meson cloud carries a charge that exceeds the proton charge, the quantity $(1-a_/2)$ is negative and hence cannot be naively associated with the ``core" size. In the SS model, this is seen in that the next-lying vector meson $\rho^\prime$ mediates, with a big coefficient, the photon-nucleon coupling with the sign opposite to that of the $\rho$. This feature which is generic in the SS model applying both to the pion and to the baryon, seems to be supported by the recent experiment of the decay $\tau^-\rightarrow K^-\pi^-K^+\nu_\tau$~\cite{taudecay}. It should be noted, however, that it is at odds with the GVDM of the Lomon type~\cite{lomon} where the $\rho^\prime$ contributes negligibly and with positive sign.

Since the formula (\ref{GE form}) does give correct charge and
correct radius, there seems to be nothing unphysical about the structure at
least at low momentum transfer: It gives as good a description of the
proton structure as (or even superior to) the QCD-motivated models like the one of \cite{bijker}. So the question is where in the holographic
description is the core ``seen" in QCD-motivated models and also in experiments? Answering this questions will require a more sophisticated analysis that we relegate to a future publication.

It is interesting to note that the infinite-tower form factor calculated in \cite{HRYY,HRYY:VD} gives also a consistent value of the $\rho NN$ coupling that in the past served as a support for the chiral bag model,
\be
\frac{\kappa_\rho}{\kappa_V}\approx 1.67
\ee
close to the empirical value
$\frac{\kappa_\rho}{\kappa_V}|_{exp}\approx 1.78$. This suggests that
the two-component model such as the chiral bag at the magic angle is
not necessarily the only compelling mechanism for the anomalous $\rho
NN$ tensor coupling.

\section{Further Comments}

While the infinite tower structure of hQCD model -- with $(a,z)\simeq
(2.44,0.08)$ -- gives the pion form factor that deviates little from
the sVD (Sakurai VD) with $(a,z)=(2,0)$, the nucleon form factor
deviates markedly from the sVD with the fit value giving $(a,z)\simeq
(4.4,0.4)$. This implies, in the way the form factor is parameterized,
that the $V^{(0)}=\rho,\omega$ contribution to the nucleon form factor
is greater by a factor of two than that to the pion, so the
conventional VD with the ``universality" does not hold at all. This
feature suggested by Nature is reproduced, semi-quantitatively, in the
hQCD model of Sakai and Sugimoto in that for instance, the charge
contributed from the lowest lying vector meson overshoots the one unit
of charge for the proton, which then is largely compensated by the
negative charge contribution coming from the first excited $\rho$ and
$\omega$ mesons. Understanding what this means will be one of the main tasks of
our future work.

Our main finding is that our two-parameter formulae inferred from the infinite tower structure of hQCD work surprisingly well for $G_{E,M}^p$ for momentum transfers $Q^2\lsim 0.5$ GeV$^2$. They are much simpler than other QCD-motivated models and provide better $\chi^2$'s. However translated into a two-component structure, namely, ``core" plus vector-dominance, they have a very different form. Given that the parameters $(a,z)$ determined in the fit seem to represent Nature, the challenge for the theorists is to calculate the parameters from a theory that reproduce the fit parameters.

Two further remarks are in order:  One on hidden local symmetry
in general and the other on the infinite tower structure in the baryon
and nuclear sector.

Hidden local symmetry on which we heavily rely is a not-yet fully
elucidated notion. It is a flavor gauge symmetry tied to emergent
gauge symmetries in theories with a weakly coupled dual description
and pervades in other areas of physics, namely in the dynamics of
supersymmetric QCD~\cite{susy} and strongly-correlated condensed
matter systems~\cite{condensed}. Applied to dense baryonic matter for
which no model-independent theoretical tools are available with
lattice QCD incapable of handling large chemical potential, it makes
an unexpected prediction that the famous short-range repulsion between
two nucleons effective in matter-free space should get {\em strongly}
suppressed as the system approaches density-driven chiral
restoration~\cite{SLPR}.  If confirmed, this mechanism will have a
drastic consequence on the equation of states of neutron stars and
other forms of compact stars.

We have noted that the infinite-tower structure in the nucleon form
factors is somewhat less successful in the magnetic response than in
the electric response. The reason for the apparent defect of the hQCD
for $G_M$ is most likely in the importance of $1/N_c$ corrections
in the Pauli form factor $F_2$, specially the isoscalar part. This is
reflected in the magnetic moment in that
$\mu_{I=0}\sim {\cal O}(1/N_c)$ while $\mu_{I=1}\sim {\cal O}(N_c)$.
Unfortunately it is
not known how to systematically calculate $1/N_c$ corrections in hQCD
models.

The other point to make is the potential importance of the infinite
tower structure in hadronic physics in general, particulary in
multi-baryon systems. In addition to the role that they play in the
nucleon form factors, the infinite tower of vector mesons in the guise
of stack of hidden local symmetry fields are found to make the nucleon
structure as a soliton qualitatively different from the Skyrme model
with pion fields only~\cite{HRYY}. Perhaps more importantly, they are
found to bring a big improvement in describing multi-nucleon systems,
i.e., nuclei, in terms of topological solitons~\cite{sutcliffe}. In
particular, it seems highly likely that the tower with their hidden
local symmetry plays an important role in dense baryonic matter
relevant to the physics of compact stars for which reliable
model-independent theoretical tools are still lacking.

\subsection*{Acknowledgments}
This work was partially supported by the WCU project of Korean
Ministry of
Education, Science and Technology (R33-2008-000-10087-0).
M.H. was supported in part by
the Grant-in-Aid for
Nagoya University Global COE Program,
``Quest
for Fundamental Principles in the Universe: from Particles to the
Solar System and the Cosmos'',
from the Ministry of Education,
Culture, Sports, Science and Technology of
Japan (MEXT),
the JSPS Grant-in-Aid for Scientific Research (S) \#22224003
\& the JSPS Grant-in-Aid for Scientific Research
(c) \#20540262 and Grant-in-Aid for Scientific Research on Innovative
Areas (\#2104)
``Quest on New Hadrons with Variety of Flavors'' from MEXT.

\section*{Appendix}
\setcounter{equation}{0}
\renewcommand{\theequation}{A\arabic{equation}}

In this Appendix, we derive the formulae (\ref{GE form}) and
(\ref{GMform}) from the action (\ref{5dfermion1}) together with
the expansion of the 5D gauge field in Eq.~(\ref{Amuxz:2}).
In the following we normalize the wave functions $\psi_{n}$ as
\begin{equation}
\int d w \frac{1}{e^2(w)}\,\psi_{(n)}(w) \psi_{(n')}(w) =
  \delta_{n n'}
\end{equation}
and introduce the mass parameter as
\begin{eqnarray}
m_n^2 \equiv - \int d w \, \frac{1}{e^2(w)}\,
  \left( \dot{\psi}_{(n)}(w) \right)^2
\ .
\end{eqnarray}
Following Ref.~\cite{HRYY}, we expand the baryon field as
\begin{equation}
{\mathcal B} = \left(
  f_L(w) \, \frac{1+\gamma_5}{2}
  + f_R(w) \, \frac{1-\gamma_5}{2}
\right) \, B
\ ,
\end{equation}
with $f_L(w) = f_R(-w)$.

We start with the equations of motion for the flavor non-singlet
vector fields $\rho_\mu^{(k)\,a} \equiv \zeta_k V_\mu^{(k) \, a}$
($k\geq 1$, $a = 1,2,\ldots,N_f^2-1$) given by
\begin{eqnarray}
0 &=&
- \int dw \frac{1}{2e^2(w)}\,\mbox{tr}\,
\left[ T_a F^{\mu 5} \right] \, \dot{\psi}_{2k+1}
\label{6 line 1}
\\
&&
{}+ \int dw \frac{1}{e^2(w)}\,\mbox{tr}\,
\left[ T_a \left\{
  \partial_\nu F^{\nu\mu} - \left[ A_\nu \,,\, F^{\nu\mu} \right]
\right\} \right] \, \psi_{(2k+1)}
\label{6 line 2}
\\
&&
{}- \int d w \bar{\mathcal B} \gamma^\mu T_a {\mathcal B}
  \, \psi_{(2k+1)}
\label{6 line 3}
\\
&&
{}+ 2 \int d w \, \kappa(w)
  \bar{\mathcal B}\gamma^{5 \mu} T_a {\mathcal B} \,\dot{\psi}_{(2k+1)}
\label{6 line 4}
\\
&&
{}- 2 \int d w \, \kappa(w)
  \psi_{(2k+1)}
 \left[
    \partial_\nu \bar{\mathcal B} \gamma^{\nu\mu} T_a {\mathcal B}
    - i \bar{\mathcal B} \gamma^{\nu\mu}
      \left[ T_a \,,\, A_\nu \right] {\mathcal B}
 \right]
\ .
\label{6 line 5}
\end{eqnarray}
Dropping the second line (\ref{6 line 2}) and the second term in
the fifth line (\ref{6 line 5}), both of which are of higher order in the chiral counting,
we obtain
\begin{eqnarray}
0 &=&
- m_{2k+1}^2 \left(
 \rho_\mu^{(k)} - \zeta_k \alpha_{\parallel\mu}
\right)
- \left( g_{V,min}^{(k)} + g_{V,mag}^{(k)} \right)
  \, \bar{B} \gamma^\mu T_a B
\nonumber\\
&&
{}- 2 \frac{g_2^{(k)}}{4m_N}\,
  \partial_\nu \left( \bar{B} \gamma^{\nu\mu} T_a B \right)
\ .
\end{eqnarray}
The quantities $g_{V,min}^{(k)}$, $g_{V,mag}^{(k)}$ and $g_2^{(k)}$
are given as $w$ integrals over the functions $f_{L,R}$ and
$\psi_{2k+1}$ as defined in (\ref{gvmin})-(\ref{g2k}).
Similarly,
the EoM for flavor-singlet vector is reduced to
\begin{eqnarray}
0 &=&
- m_{2k+1}^2 \, \left(
  \rho^{(k) \mu a =0}
  - \zeta_k {\alpha}_{\parallel}^{\mu a=0}
\right)
- g_{V,min}^{(k)} \, \bar{B} \gamma^\mu T_0 B
\ .
\end{eqnarray}
The EoM for axial-vector mesons ($m \geq 1$) becomes
\begin{eqnarray}
0 &=&
m_{2m}^2 \, A^{(k) \mu a}
- \left( g_{A,min}^{(m)} + g_{A,mag}^{(m)} \right)
  \, \bar{B} \gamma^\mu \gamma_5 T_a B
\ .
\end{eqnarray}

By substituting the above EoMs, the 5D gauge field after
integrating out becomes\footnote{The axial vector fields are not
  needed for our purpose but they are kept for completeness.}
\begin{eqnarray}
\lefteqn{
  A_\mu^{\rm integ}(x, w)
}
\nonumber\\
 &=& \alpha_{\parallel\mu} (x)
  -
  \hat{\alpha}_{\perp\mu}(x) \psi_{(0)}(w)
\nonumber \\
&&
{}+
\sum_{k=1}^\infty
\frac{g_{A,min}^{(k)} + g_{A,mag}^{(k)}}{m_{2k}^2}
\, \bar{B} \gamma^\mu \gamma_5 T_a B
\, \psi_{(2m)}(w)
\, T_a
\nonumber\\
&&
{}+
\left(
  \rho_\mu^{(0)} - \zeta_0 \alpha_{\parallel \mu}
\right)
\psi_{(1)}(w)
\nonumber\\
&&
{}-
\sum_{k=1}^\infty \sum_{a=0}^{3}
\frac{1}{m_{2k+1}^2}
\left[ \widetilde{g}_V^{(k)}
  \, \bar{B} \gamma_\mu T_a B
+ 2 \frac{g_2^{(k)}}{4m_N}\,
  \partial^\nu \left( \bar{B} \gamma_{\nu\mu} T_a B \right)
\right]
\psi_{(2k+1)}(w)
\, T_a
\nonumber\\
&&
{} -
\sum_{k=1}^\infty
\frac{g_{V,min}^{(k)}}{m_{2k+1}^2}
  \, \bar{B} \gamma_\mu T_0 B
\psi_{(2k+1)}(w) \, T_0
\, ,
\label{Amuxz:3}
\end{eqnarray}
where
\begin{equation}
\widetilde{g}_V^{(n)} \equiv
g_{V,min}^{(n)} + g_{V,mag}^{(n)}
\ .
\end{equation}
Using this $A_\mu^{\rm integ}$, the field strengths are
obtained as
\begin{eqnarray}
\lefteqn{
  F_{5\mu}^{\rm integ}(x, w)
}
\nonumber\\
&=&
- \hat{\alpha}_{\perp\mu} \partial_w \psi_{(0)}
+ \left(
  \rho_\mu^{(0)} - \zeta_0 \alpha_{\parallel \mu}
\right)  \partial_w \psi_{(1)}
\nonumber\\
&&
{}-
\sum_{k=1}^\infty \sum_{a=0}^{3}
\frac{1}{m_{2k+1}^2}
\left[ \widetilde{g}_V^{(k)}
  \, \bar{B} \gamma_\mu T_a B
+ 2 \frac{g_2^{(k)}}{4m_N}\,
  \partial^\nu \left( \bar{B} \gamma_{\nu\mu} T_a B \right)
\right]
\partial_w \psi_{(2k+1)}
\, T_a
\nonumber\\
&&
{} -
\sum_{k=1}^\infty
\frac{g_{V,min}^{(k)}}{m_{2k+1}^2}
  \, \bar{B} \gamma_\mu T_0 B
\partial_w \psi_{(2k+1)} \, T_0
\, ,
\\
\lefteqn{
  F_{\mu\nu}^{\rm integ}(x, w)
}
\nonumber\\
&=&
\partial_\mu \alpha_{\parallel \nu} -
\partial_\mu \hat{\alpha}_{\perp \nu} \,\psi
\nonumber\\
&&
{}+
\left(
  \rho_\nu^{(0)} - \zeta_0 \alpha_{\parallel \nu}
\right)
\psi_{(1)}
\nonumber\\
&&
{}-
\sum_{k=1}^\infty \sum_{a=0}^{3}
\frac{1}{m_{2k+1}^2}
\partial_\mu
\left[ \widetilde{g}_V^{(k)}
  \, \bar{B} \gamma_\nu T_a B
+ 2 \frac{g_2^{(k)}}{4m_N}\,
  \partial^\sigma \left( \bar{B} \gamma_{\sigma\nu} T_a B \right)
\right]
\psi_{(2k+1)}(w)
\, T_a
\nonumber\\
&&
{} -
\sum_{k=1}^\infty
\frac{g_{V,min}^{(k)}}{m_{2k+1}^2}
  \, \partial_\mu \bar{B} \gamma_\nu T_0 B
\psi_{(2k+1)}(w) \, T_0
\nonumber\\
&&
{}- \,\left( \mu \leftrightarrow \nu \right)
\ \cdots
\ .
\end{eqnarray}
{}From this, the relevant terms of action become
\begin{eqnarray}
S &=&
\int d^4x d w
\Biggl[
  - \overline{\mathcal B} \gamma^\mu A_{\mu}^{\rm integ} {\mathcal B}
 + 2 \kappa(w) \overline{\mathcal B} \gamma_5 \gamma^\mu
    F_{5\mu}^{\rm integ; SU(2)}{\mathcal B}
 + \kappa(w) \overline{\mathcal B} \gamma^{\mu\nu}
    F_{\mu\nu}^{\rm integ; SU(2)} {\mathcal B}
\nonumber\\
&&
- \frac{1}{2 e^2(w)}
\left\{
  2 \mbox{tr} \left(
    F_{5\mu}^{\rm integ} F^{5\mu}_{\rm integ}
  \right)
  + \mbox{tr} \left(
    F_{\mu\nu}^{\rm integ} F^{\mu\nu}_{\rm integ}
  \right)
\right\}
\Biggr]
\nonumber\\
&=&
\int d^4 x
\Biggl[
 -\bar{B}\gamma^\mu
 \left\{
   \alpha_{\parallel\mu}
   +
   g_{V,min}^{(0)}
  \left(
     \rho_\nu^{(0)} -\zeta_0 \alpha_{\parallel\mu}
   \right)
 \right\}
 {B}
{}- g_{V,mag}^{(0)}
\bar{B}\gamma^{\mu}
\left(
  \rho_\nu^{(0)} - \zeta_0 \alpha_{\parallel \nu}
\right)
\,{B}
\nonumber\\
&& \qquad
{}+ 2
\bar{B}\gamma^{\mu\nu}
  \left\{
    \frac{g_2}{4m_N}
      \partial_\mu \alpha_{\parallel\nu}^{\rm SU(2)} (x)
    + \frac{g_2^{(0)}}{4m_N}
      \partial_\mu \left(
         \rho_\nu^{(0)} - \zeta_0 \alpha_{\parallel \nu}
      \right)
  \right\}
{B}
\Biggr]
\nonumber\\
&& {}- 2\int d^4 x  \,
\mbox{tr}\,
\Biggl[
\biggl\{
 \partial_\nu \alpha_{\parallel\mu} (x)
 - \partial_\mu \alpha_{\parallel\nu} (x)
\biggr\}
\nonumber\\
&& \quad
\times \biggl\{
 - \partial_\nu
  \sum_{n=1}^\infty
  \frac{\zeta_n}{m_{2n+1}^2}\,
  \left[ \widetilde{g}_V^{(n)}
    \, \bar{B} \gamma^\mu T_a B
  + 2 \frac{g_2^{(n)}}{4m_N}\,
    \partial_\sigma \left( \bar{B} \gamma^{\sigma\mu} T_a B \right)
  \right]
  \, T_a
\nonumber\\
&& \qquad
  {}-
  \partial_\nu
  \sum_{n=1}^\infty
  \frac{\zeta_n}{m_{2n+1}^2}\,
  g_{V,min}^{(n)}
    \, \bar{B} \gamma^\mu T_0 B
  \, T_0
\biggr\}
\Biggr]
\ + \cdots
\ .
\label{reduced 1}
\end{eqnarray}
The field $\alpha_{\parallel \mu}$ is expanded as
\begin{equation}
\alpha_{\parallel \mu} = \tilde{A}_\mu Q \,+\,\cdots
\ ,
\end{equation}
where
$\tilde{A}_\mu$ is the photon field and
$Q = \mbox{diag.}(1\,,\,0)$ is the charge matrix.
Substituting this, we obtain
\begin{eqnarray}
S &=&
\int d^4 x
\Biggl[
 -\bar{B}\gamma^\mu
 \tilde{A}_\mu
 \left\{
   Q \left( 1 - g_{V,min}^{(0)} \zeta_0 \right)
   + \frac{\tau_3}{2} g_{V,mag}^{(0)} \zeta_0
 \right\}
 {B}
\nonumber\\
&& \
- \bar{B} \gamma^\mu
 \left\{
   g_{V,min}^{(0)} \rho_\mu
   + g_{V,mag}^{(0)} \rho_\mu^{SU(2)}
 \right\}
 {B}
\nonumber\\
&& \
+ \frac{1}{2m_N} \bar{B} \gamma^{\mu\nu} \frac{\tau_3}{2}
  \partial_\mu \tilde{A}_\nu
 \left(
   g_2 - g_2^{(0)} \zeta_0
 \right)
- g_2^{(0)} \bar{B}\gamma^{\mu\nu}
  \partial_\mu \rho_\nu^{SU(2)}
 B
\nonumber\\
&& \
  - \int d^4x
\bar{B}\Biggl[
  \sum_{k=1}^{\infty}
  \frac{\zeta_k}{m_{2k+1}^2}
  \gamma^\mu \partial_\sigma \partial^\sigma \tilde{A}_\mu
  \left( \tilde{g}_{V}^{(k)} Q + g_{V,min}^{(k)} T_0 \right)
\nonumber\\
&& \quad
  - \sum_{k=1}^\infty \frac{\zeta_k}{m_{2k+1}^2}
  \frac{g_2^{(k)}}{2m_N}
  \gamma^{\mu\nu} \partial_\sigma \partial^\sigma
  \partial_\nu \tilde{A} Q
\Biggr] B
\ .
\label{reduced 3}
\end{eqnarray}
Now we can read off the Dirac and Pauli form factors from
Eq.(\ref{reduced 3}):
\begin{eqnarray}
F_1^p(Q^2) &=&
  \left(
    g_{V,{\rm min}}^{(0)} + \frac{1}{2} g_{V,{\rm mag}}^{(0)}
  \right)
  \frac{\zeta_0 \, m_{V}^2 }
   {Q^2 + m_{V}^2}
\nonumber\\
&&
  {} +
  \sum_{k=1}^{\infty}
  \left(
    g_{V,{\rm min}}^{(k)} + \frac{1}{2} g_{V,{\rm mag}}^{(k)}
  \right)
  \zeta_k \,
  \left[ 1 - \frac{Q^2}{m_{2k+1}^2} \right]
\ ,
\nonumber\\
F_2^p(Q^2) &=&
\frac{1}{2}\,\frac{g_2^{(0)} \, \zeta_0 m_V^2 }
   {Q^2 + m_{V}^2}
+ \frac{1}{2} \left( g_2 - g_2^{(0)} \zeta_0 \right)
- \frac{1}{2}\,\sum_{k=1}^{\infty}
  g_2^{(k)}\zeta_k\,\frac{Q^2}{m_{2k+1}^2}
\ .
\end{eqnarray}
The Sachs form factors follow from the definitions (\ref{GE}) and
(\ref{GM}).


\end{document}